\documentclass[aps,twocolumn,a4paper,nofootinbib,superscriptaddress]{revtex4}
\usepackage[dvips]{graphicx}
\usepackage{amsmath}
\usepackage{amssymb}

\begin{document}

\title{Many-body renormalization of Landau levels in graphene due to screened Coulomb interaction}
\author{Alexey A. Sokolik}%
\affiliation{Institute for Spectroscopy, Russian Academy of Sciences, 142190 Troitsk, Moscow, Russia}%
\affiliation{National Research University Higher School of Economics, 109028 Moscow, Russia}%
\author{Yurii E. Lozovik}%
\email{lozovik@isan.troitsk.ru}
\affiliation{Institute for Spectroscopy, Russian Academy of Sciences, 142190 Troitsk, Moscow, Russia}%
\affiliation{National Research University Higher School of Economics, 109028 Moscow, Russia}
\affiliation{Dukhov Research Institute of Automatics (VNIIA), 127055 Moscow, Russia}%

\begin{abstract}
Renormalization of Landau level energies in graphene in strong magnetic field due to Coulomb interaction is studied
theoretically, and calculations are compared with two experiments on carrier-density dependent scanning tunneling
spectroscopy. An approximate preservation of the square-root dependence of the energies of Landau levels on their
numbers and magnetic field in the presence of the interaction is examined. Many-body calculations of the renormalized
Fermi velocity with the statically screened interaction taken in the random-phase approximation show good agreement
with both experiments. The crucial role of the screening in achieving quantitative agreement is found. The main
contribution to the observed rapid logarithmic growth of the renormalized Fermi velocity on approach to the charge
neutrality point turned out to be caused not by mere exchange interaction effects, but by weakening of the screening at
decreasing carrier density. The importance of a self-consistent treatment of the screening is also demonstrated.
\end{abstract}

\maketitle

\section{Introduction}

Many-body effects of Coulomb interaction between massless Dirac electrons in graphene are widely studied both
theoretically and in a series of transport and optical experiments \cite{CastroNeto,Kotov,Basov}. Renormalization of
the electron Fermi velocity to higher values, showing logarithmical divergence upon approaching the charge neutrality
point \cite{Kotov,Gonzalez}, is the most prominent signature.

Graphene in quantizing magnetic field serves as a perspective system where both quantum single-particle and many-body
effects can be studied \cite{Goerbig,Orlita,Miransky,Yin1}. According to an idealized single-particle picture,
electrons in monolayer graphene in the magnetic field $B$ occupy the relativistic Landau levels
\begin{eqnarray}
E_n=\mathrm{sgn}(n)\,v_\mathrm{F}\sqrt{2|n|Be\hbar/c},\quad n=0,\pm1,\pm2,\ldots,\label{E_n}
\end{eqnarray}
where $v_\mathrm{F}\approx10^6\,\mbox{m/s}$ is the Fermi velocity. In the presence of Coulomb interaction, many-body
effects cause renormalization of these energies to new values $E_n^*$. Thus the problem of systematization and
theoretical description of the interaction induced energy shifts $E_n^*-E_n$ arises.

In several experiments \cite{Li1,Li2,Luican_SSC,Miller,Song,Luican,Chae,Yin2} on scanning tunneling spectroscopy of
graphene in magnetic field, $E_n^*$ were measured and turned out to be in agreement with the same square-root law
$E_n^*\propto\mathrm{sgn}(n)\sqrt{|n|B}$ as for noninteracting electrons. This admits an effective single-particle
description of the energy levels in a many-body system using the phenomenological renormalized Fermi velocity
$v_\mathrm{F}^*$:
\begin{eqnarray}
E_n^*=\mathrm{sgn}(n)\,v_\mathrm{F}^*\sqrt{2|n|Be\hbar/c}.\label{E_n_ren}
\end{eqnarray}

However some of these experiments \cite{Luican,Chae}, where the carrier density in graphene $n_\mathrm{e}$ was varied,
reported the growth of $v_\mathrm{F}^*$ when the density of electrons or holes decreases approaching the charge
neutrality point $n_\mathrm{e}=0$. This effect cannot be described by an effective single-particle picture with a fixed
$v_\mathrm{F}^*$. It has intrinsically many-body character and bears similarities to renormalization of the effective
Fermi velocity predicted and observed in the absence of magnetic field \cite{Kotov,Gonzalez}.

Deviations of observable energies of inter-Landau level transitions from single-particle theory predictions, discovered
in cyclotron resonance \cite{Jiang1,Jiang2,Henriksen,Russell} and magneto-Raman scattering
\cite{Berciaud,Faugeras1,Faugeras2,Sonntag} experiments, are another manifestation of many-body effects in graphene in
magnetic field. Proper theoretical description of these deviations requires taking into account not only
renormalization of individual Landau levels, but also the excitonic effects of electron-hole interaction in a final
state \cite{Iyengar,Bychkov,Barlas,Roldan1,Shizuya1,Shizuya2,Faugeras1,Sokolik,Shizuya4,Sonntag}. The splitting of
Landau levels in very strong magnetic field can be also mentioned as a striking signature of many-body effects (see,
e.g., \cite{Goerbig,Miransky} and references therein).

In this paper, we focus our theoretical study on renormalization of Landau levels in monolayer graphene, which was
observed in the scanning tunneling spectroscopy experiments \cite{Luican,Chae} where the variations of $v_\mathrm{F}^*$
versus $n_\mathrm{e}$ were measured. Our calculations are based on evaluating the mean-field exchange energies of
electrons at Landau levels using the statically screened Coulomb interaction, as described in Sec.~\ref{sec2}.

Then in Sec.~\ref{sec3} we analyze the behavior of the renormalized energies and discuss the validity of approximating
a set of $E_n^*$ with the single formula (\ref{E_n_ren}), which is routinely applied in the experimental works
\cite{Li1,Li2,Luican_SSC,Miller,Song,Luican,Chae,Yin2}.

In Sec.~\ref{sec4} we fit the experimental data with our calculations in different approximations, both with and
without screening. The bare Fermi velocity $v_\mathrm{F}=0.85\times10^6\,\mbox{m/s}$ and the realistic dielectric
constants allow us to achieve good agreement with the experiments. We show that the main cause of the renormalized
Fermi velocity growth at decreasing carrier density is not the Landau-level filling factor dependence of electron
exchange energies, but the weakening of the screening. Therefore taking into account the interaction screening is
crucial for appropriate description of the many-body effects. The important role of the self-consistent weakening of
the screening is also demonstrated. Our conclusions are presented in Sec.~\ref{sec5}.

\section{Theoretical model}\label{sec2}

The conventional approach
\cite{Iyengar,Bychkov,Roldan1,Barlas,Shizuya1,Shizuya2,Gorbar,Chizhova,Menezes,Shovkovy,Yan,Shizuya4,Sokolik} to treat
the interaction-induced renormalization of Landau level energies is based on the Hartree-Fock approximation, where the
renormalized energy
\begin{eqnarray}
E_n^*=E_n+\Sigma_n-\Sigma_0\label{E_n_ren_calc}
\end{eqnarray}
consists of the bare single-particle energy $E_n$ and of the exchange self-energy $\Sigma_n$. To maintain the Dirac
point location at $E=0$, the renormalized energy $E_0^*=\Sigma_0$ of the zeroth Landau level was subtracted. The
self-energy
\begin{eqnarray}
\Sigma_n=-\sum_{n'k'}f_{n'}\langle\psi_{nk},\psi_{n'k'}|V|\psi_{nk},\psi_{n'k'}\rangle\label{Sigma}
\end{eqnarray}
is a result of virtual processes of exchanging an electron on the $n$th level with other electrons in all occupied
$n'$th levels; $f_{n'}$ are the occupation numbers of these levels ($0\leqslant f_{n'}\leqslant1$), and
$\langle\psi_{nk},\psi_{n'k'}|V|\psi_{nk},\psi_{n'k'}\rangle$ are the exchange matrix elements of the Coulomb
interaction $V(\mathbf{r})$. The single-particle states $\psi_{nk}$ are characterized by the Landau level number $n$
and by an additional quantum number $k$, which accounts for a Landau level degeneracy.

To handle the logarithmic divergence of the self-energies (\ref{Sigma}) upon summation over the negative-energy Landau
levels $n'$, the physically motivated cutoff $n'\geqslant-n_\mathrm{c}$ should be introduced
\cite{Iyengar,Bychkov,Roldan1,Barlas,Shizuya1,Shizuya2}. Because of its degeneracy, each Landau level hosts
$geB/2\pi\hbar c$ electrons per unit area, where $g=4$ is the fourfold spin and valley degeneracy factor. Assuming that
$n_\mathrm{c}$ occupied Landau levels below the Dirac point in charge neutral graphene should host two electrons per
elementary cell of the area $S_0=a^2\sqrt3/2$ ($a\approx2.46\,\mbox{\AA}$), we get
\begin{eqnarray}
n_\mathrm{c}=\frac{8\pi\hbar c}{\sqrt3ga^2eB}\approx\frac{39\,600}{B\,\mbox{[T]}}.\label{n_c}
\end{eqnarray}
As noted is \cite{Yan}, a numerical calculation and summation of large numbers, typically tens of thousands, of matrix
elements in (\ref{Sigma}) presents a separate computational problem. The formula (\ref{n_c}) contains some
uncertainties due to inaccuracy of the Dirac low-energy model away from the Dirac point, but our estimates show their
influence of the calculation results is weak: for example, changing the number $39\,600$ in the numerator of
(\ref{n_c}) by 20\% results in the change of the calculated $E_n^*$ by about 1\%.

In the most of theoretical works \cite{Iyengar,Bychkov,Barlas,Roldan1,Shizuya1,Shizuya2,Shizuya4,Chizhova,Menezes}, the
unscreened Coulomb interaction $v(r)=e^2/\varepsilon r$ was used to calculate the exchange energies (\ref{Sigma}). Here
$\varepsilon$ is the dielectric constant of a surrounding medium, which varies from $\varepsilon=1$ for suspended
graphene to $\varepsilon\sim10-20$ for graphene on a graphite substrate \cite{Faugeras1,Sokolik}. In our earlier paper
\cite{Sokolik} we showed that the screening of Coulomb interaction by a polarizable gas of massless electrons of
graphene should be taken into account to achieve quantitative agreement with the experiments on cyclotron resonance and
magneto-Raman scattering \cite{Jiang1,Jiang2,Berciaud,Faugeras1}. In this work we follow the similar approach and use
the statically screened interaction when calculate the matrix elements in (\ref{Sigma}). Its Fourier transform is
\begin{eqnarray}
V(q)=\frac{v(q)}{1-v(q)\Pi(q,0)},\label{V_scr}
\end{eqnarray}
where $v(q)=2\pi e^2/\varepsilon q$ is the Fourier transform of the unscreened interaction $v(r)$, $\Pi(q,\omega)$ is
the irreducible polarizability of graphene in magnetic field. In the random-phase approximation, the latter is
calculated by taking into account all virtual electron transitions between filled and empty Landau levels,
\begin{eqnarray}
\Pi(q,0)=g\sum_{nn'}F_{nn'}(q)\frac{f_n-f_{n'}}{E_n-E_{n'}},\label{Pol}
\end{eqnarray}
where $F_{nn'}(q)$ are the Landau level form factors (see the details in
\cite{Goerbig,Gumbs,Pyatkovskiy,Roldan2,Roldan3,Tahir}). Note that the statically screened interaction was also used in
\cite{Gorbar,Shovkovy,Yan} to analyze possible spontaneous symmetry breaking and gap generation scenarios. Generally,
full dynamical treatment of the screening can provide more accurate results than in the static approximation, although
the computational procedure in this case becomes very demanding.

Let us examine how the screened interaction $V(q)$ depends on the parameters $v_\mathrm{F}$, $\varepsilon$ of the
model. Via the energy denominators, the polarizability (\ref{Pol}) scales with the Fermi velocity as $\Pi(q,0)\propto
v_\mathrm{F}^{-1}$, thus the quantity $v(q)\Pi(q,0)$ in the denominator of (\ref{V_scr}) scales as
\begin{eqnarray}
v(q)\Pi(q,0)\propto r_\mathrm{s}\equiv\frac{e^2}{\varepsilon\hbar v_\mathrm{F}}.
\end{eqnarray}
The dimensionless parameter $r_\mathrm{s}$ (sometimes referred to as the graphene analogue $\alpha$ of the fine
structure constant \cite{Kotov,Goerbig}) is conventionally used to characterize the relative strength of Coulomb
interaction. However in our approach it actually measures the strength of the screening due to the linear scaling
between $r_\mathrm{s}$ and $\Pi(q,0)$ in (\ref{V_scr}).

Another important point is a necessity of a self-consistent treatment of the screening, i.e. calculation of the
polarizability using the electron Green functions, which are already ``dressed'' by many-body interaction effects. In
the simplest approach, this reduces to a calculation of the energy differences in the denominators of (\ref{Pol}) with
the renormalized energies $E_n^*$ instead of the bare ones $E_n$. Since the typical renormalized Fermi velocities
$v_\mathrm{F}^*$ can be up to $60\%$ higher than the bare velocity $v_\mathrm{F}$ \cite{Sokolik} (and the same is true
for the energy differences), the self consistent screening can be appreciably weaker than in the random-phase
approximation. Full self-consistent treatment of the screening is highly computationally demanding so we use the
simplified model where $\Pi(q,0)$ retains its functional form but becomes proportionally reduced by the factor
$v_\mathrm{F}/v_\mathrm{F}^*<1$. This is achieved by replacing $v_\mathrm{F}$ in the definition of $r_\mathrm{s}$ by
$v_\mathrm{F}^*$, so the resulting $r_\mathrm{s}$, which appears in the denominator of (\ref{V_scr}), becomes smaller
than in the case of the non self-consistent screening. The value of $v_\mathrm{F}^*$ in the prefactor of $\Pi(q,0)$ can
be taken either from experiments or from theoretical calculations.

The assumption of the unchanged functional form of $\Pi(q,0)$ is reasonable because the renormalized Landau level
energies retain their dependence $E_n^*\propto\mathrm{sgn}(n)\sqrt{|n|B}$ on $n$, $B$ with sufficiently high accuracy
(as confirmed in the next section) and only change by the overall numerical factor $v_\mathrm{F}^*/v_\mathrm{F}$. The
remote levels with with $|n|\gg1$ can deviate from this regularity because of a breakdown of the Dirac model at high
energies, but their contribution to $\Pi(q,0)$ is not critical due to small form factors $F_{nn'}(q)$ and large energy
denominators.

In order to analyze the role of the screening, we carry out the calculations using, similarly to \cite{Sokolik}, the
following four screening models:

(1) unscreened Coulomb interaction, when $r_\mathrm{s}=0$;

(2) screened interaction, $r_\mathrm{s}=e^2/\varepsilon\hbar v_\mathrm{F}$;

(3) self-consistently screened interaction with $r_\mathrm{s}=e^2/\varepsilon\hbar\langle v_\mathrm{F}^*\rangle$, where
$\langle v_\mathrm{F}^*\rangle$ is the average value of the experimental renormalized Fermi velocity in the measured
range of carrier densities;

(4) self-consistently screened interaction with a variable screening strength $r_\mathrm{s}=e^2/\varepsilon\hbar
v_\mathrm{F}^*(n_\mathrm{e})$ where $r_\mathrm{s}$ and $v_\mathrm{F}^*(n_\mathrm{e})$ are calculated at each
$n_\mathrm{e}$ iteratively: initially $v_\mathrm{F}^*=v_\mathrm{F}$ and then in each iteration the next value of
$v_\mathrm{F}^*$ is extracted from a set of $E_n^*$ calculated with $r_\mathrm{s}=e^2/\varepsilon\hbar v_\mathrm{F}^*$
determined by the previous value of $v_\mathrm{F}^*$ (typically 5 iterations are sufficient to achieve convergence).

The 3rd and 4th models consider the screening to be self-consistently weakened with respect to that in the random-phase
approximation: in the 3rd model the magnitude of this weakening is estimated on the basis of experimental data, while
in the 4th model it is calculated theoretically. We will show below that the 3rd and 4th models provide much better
agreement with the experimental data indicating the importance of the self-consistent treatment of the screening.

\begin{figure}[t]
\begin{center}
\resizebox{0.9\columnwidth}{!}{\includegraphics{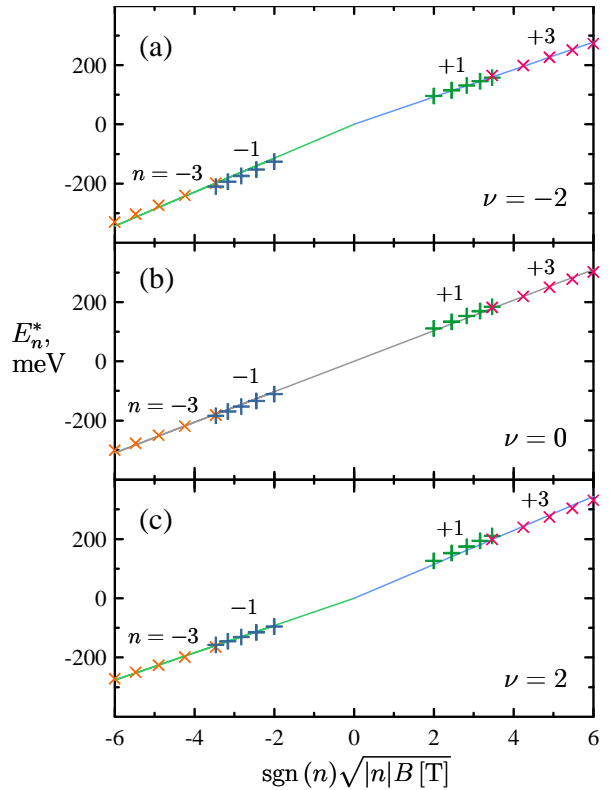}}
\end{center}
\caption{\label{Fig1}Renormalized Landau level energies $E_n^*$ as functions of $\mathrm{sgn}(n)\sqrt{|n|B}$ calculated
with $v_\mathrm{F}=0.85\times10^6\,\mbox{m/s}$, $\varepsilon=2$, $r_\mathrm{s}=0.87$ at different filling factors
$\nu$. Each level with the number $n=-3$, $-1$, $+1$, or $+3$ is presented by the series of five points (crosses) at
magnetic fields $B=4,6,8,10,12\,\mbox{T}$. The linear fits for these sets of points on the electron and hole sides are
shown by the solid lines.}
\end{figure}

\section{Evaluation of $v_\mathrm{F}^*$}\label{sec3}

The renormalized Landau level energies (\ref{E_n_ren_calc}) depend on the Landau level number $n$, magnetic field $B$
and, importantly, on the filling factor $\nu$. The latter equals to 0 in undoped graphene, when the zeroth level is
half-filled, while a complete filling of each new Landau level increases $\nu$ by 4 because of the fourfold spin-valley
degeneracy of electron states in graphene \cite{Goerbig}. The filling factor is related to the carrier density as
\begin{eqnarray}
n_\mathrm{e}=\frac{eB}{2\pi\hbar c}\,\nu\approx0.0242\,\nu\times10^{12}\,\mbox{cm}^{-2}\times B\,\mbox{[T]}\label{n_e}
\end{eqnarray}
and reaches the values up to $\nu\approx\pm(15-30)$ in the experiments \cite{Luican,Chae}.

It was commonly accepted \cite{Li1,Li2,Luican_SSC,Miller,Song,Luican,Chae,Yin2} that
$E_n^*\propto\mathrm{sgn}(n)\sqrt{|n|B}$ at fixed $n_\mathrm{e}$ and varying $(n,B)$. However the recent experiments
\cite{Berciaud,Faugeras1,Faugeras2} on magneto-Raman scattering revealed the visible dependence of $v_\mathrm{F}^*$ on
a magnetic field of the form $v_\mathrm{F}^*\approx C_1-C_2\ln B$. Its origin can be traced theoretically from
Eq.~(\ref{Sigma}): as the sum has a logarithmic divergence which have been regularized using the cutoff (\ref{n_c}), so
the result will be logarithmically dependent on a cutoff position and hence on $B$.

In Fig.~\ref{Fig1}, the typical dependence of $E_n^*$ on $n$, $B$, and $\nu$ is investigated. The calculations are
carried out under the typical conditions of the experiment \cite{Luican} in the 3rd aforementioned screening model (see
Sec.~\ref{sec3}). Analyzing behavior of the linear fits (\ref{E_n_ren}) as well as deviations of individual points from
these fits, we can note the following regularities:

\begin{figure}[t]
\begin{center}
\resizebox{0.9\columnwidth}{!}{\includegraphics{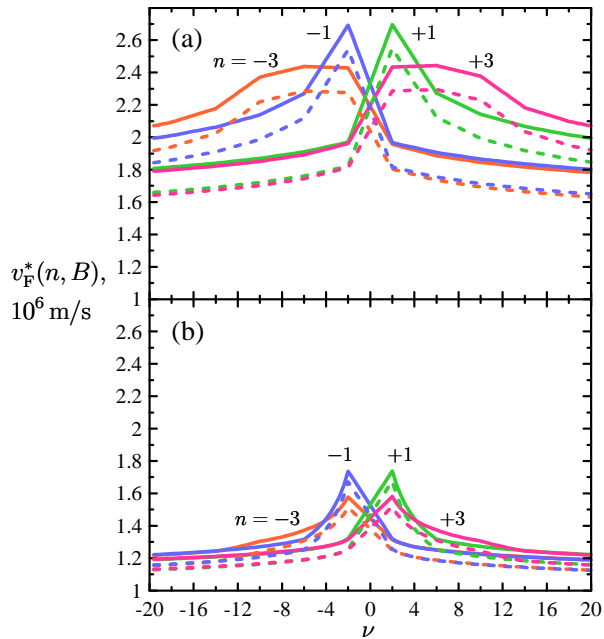}}
\end{center}
\caption{\label{Fig2}Effective Fermi velocities $v_\mathrm{F}^*(n,B)$ [see Eq.~(\ref{vF_eff})] as functions of the
filling factor $\nu$, calculated at $v_\mathrm{F}=0.85\times10^6\,\mbox{m/s}$, $\varepsilon=2$ (a) with the unscreened
interaction, and (b) with the self-consistently screened interaction ($r_\mathrm{s}=0.87$). Solid and dashed lines
correspond, respectively, to magnetic fields $B=4$ and $12\,\mbox{T}$.}
\end{figure}

(a) The effective Fermi velocities
\begin{eqnarray}
v_\mathrm{F}^*(n,B)=\frac{E_n^*}{\mathrm{sgn}(n)\sqrt{2|n|Be\hbar/c}}\label{vF_eff}
\end{eqnarray}
assigned to individual points $(n,B)$ always decrease at increasing $B$. For example, the $n=+3$ points in
Fig.~\ref{Fig1}(a) lie slightly above the linear fit at low $B$ and slightly below the fit at high $B$. This conforms
both the experiments and the theory demonstrating the logarithmic decrease of $v_\mathrm{F}^*$ versus $B$
\cite{Berciaud,Faugeras1,Faugeras2,Shizuya1,Shizuya2,Sokolik,Shizuya4}.

(b) The electron-hole asymmetry between Fig.~\ref{Fig1}(a) and Fig.~\ref{Fig1}(c) implies that $v_\mathrm{F}^*(n,B)$ is
higher when $n$ and $\nu$ have the same sign (both are on the electron or on the hole side) and lower when they have
different signs (one is on the electron, and the other is on the hole side). In other words, the energies of those
Landau levels which lie closer to the Fermi level undergo stronger renormalization to higher values.

(c) The effect of the upward renormalization of the Fermi velocity is generally more pronounced for Landau levels with
smaller $|n|$. For example, in Fig.~\ref{Fig1} it is appreciably stronger for $n=\pm1$ than for $n=\pm3$.

The same regularities are presented more clearly in Fig.~\ref{Fig2}, where the effective Fermi velocities
$v_\mathrm{F}^*(n,B)$ are plotted as functions of the filling factor $\nu$. Without the screening [Fig.~\ref{Fig2}(a)],
$v_\mathrm{F}^*(n,B)$ are the piecewise-linear functions of $\nu$ with the kinks at $\nu=\pm2,\pm6,\pm10,\ldots$, where
an integer number of Landau levels is filled. This behavior is dictated by Eq.~(\ref{Sigma}), which implies a linear
change of $E_n^*$ versus $f_{n'}$ during filling of each Landau level. The screening [Fig.~\ref{Fig2}(b)] significantly
reduces the effective Fermi velocities and removes the strict piecewise-linear behavior because now the screened
interaction $V(q)$ also changes during the filling of each Landau level. Nevertheless, an approximate piecewise
linearity survives. Note that the evident piecewise linear dependence was observed in the recent magneto-Raman
scattering experiment \cite{Sonntag}.

In the scanning tunneling spectroscopy experiments \cite{Li1,Li2,Luican_SSC,Miller,Song,Luican,Chae,Yin2}, the values
of $v_\mathrm{F}^*$ were evaluated by approximating some set of measured $E_n^*$ with different $(n,B)$ by the formula
(\ref{E_n_ren}). In \cite{Luican,Chae}, this procedure was repeated at each density $n_\mathrm{e}$ do obtain the
$v_\mathrm{F}^*(n_\mathrm{e})$ relationship. The presented analysis allows us to conclude that the result of such
evaluation can depend on a taken set of $(n,B)$. Generally, higher $B$ and $|n|$ lead to lower $v_\mathrm{F}$, and the
electron-hole asymmetry can influence the result, though in the most cases the approximation (\ref{E_n_ren}) is quite
accurate.

\section{Comparison with experiments}\label{sec4}

\begin{figure}[t]
\begin{center}
\resizebox{0.9\columnwidth}{!}{\includegraphics{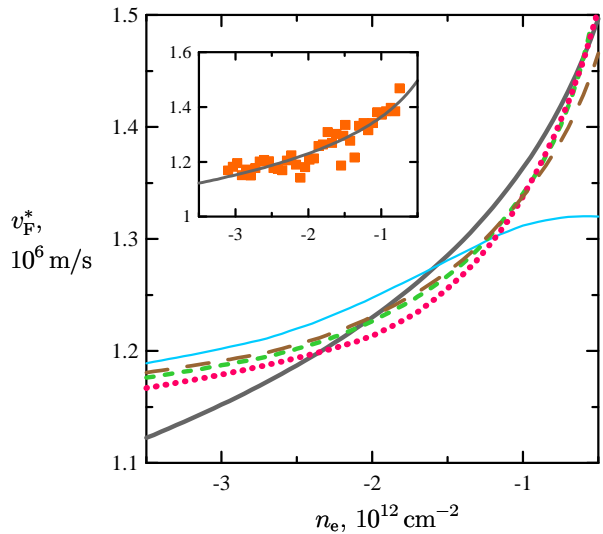}}
\end{center}
\caption{\label{Fig3}The best fits to the experimental \cite{Luican} dependence of $v_\mathrm{F}^*$ on the carrier
density $n_\mathrm{e}$ at $v_\mathrm{F}=0.85\times10^6\,\mbox{m/s}$: with the unscreened interaction (thin solid line),
screened interaction (short-dashed line), self-consistently screened interaction with the constant $r_\mathrm{s}$
(long-dashed line) and self-consistently screened interaction with the iteratively calculated varying $r_\mathrm{s}$
(dotted line). The logarithmic trend (\ref{Luican_trend}) of the experimental Fermi velocity growth is shown by think
solid line. Inset shows the experimental points (squares) and the extracted logarithmic trend.}
\end{figure}

In the first considered experiment \cite{Luican}, graphene on a $\mathrm{SiO}_2$ substrate was studied and the fitting
of the energies of several lowest Landau levels $n=-3,-2,-1,0$ at magnetic fields in the range $B=4-8\,\mbox{T}$ was
reported. The carrier densities $n_\mathrm{e}$ were taken at rather low hole doping levels not exceeding
$-3\times10^{12}\,\mbox{cm}^{-2}$. As shown in the inset in Fig.~\ref{Fig3}, $v_\mathrm{F}^*$ in spite of some
scattering of the experimental points demonstrates the clear growing trend at decreasing $|n_\mathrm{e}|$ which can be
approximated by the logarithmic function
\begin{eqnarray}
\frac{v_\mathrm{F}^*(n_\mathrm{e})}{10^6\,\mbox{m/s}}=
\left(1.363-0.192\ln\frac{|n_\mathrm{e}|}{10^{12}\,\mbox{cm}^{-2}}\right).\label{Luican_trend}
\end{eqnarray}
To reproduce these results theoretically, we have taken the set of 20 energies $E_n^*$ with $n=-3,-2,-1,0$ at
$B=4,5,6,7,8\,\mbox{T}$ and evaluated $v_\mathrm{F}^*$ at each $n_\mathrm{e}$ by their least-square fitting with the
formula (\ref{E_n_ren}). Note that even at fixed $n_\mathrm{e}$ the filling factor $\nu$ varies in the calculations
when $B$ is changed due to Eq.~(\ref{n_c}).

Each of the four calculation models, described in Sec.~\ref{sec2}, need the bare Fermi velocity $v_\mathrm{F}$ and the
dielectric constant of surroundings $\varepsilon$ as the parameters. As known (see, e.g., \cite{Sokolik,Lozovik}), the
phenomenological values of $v_\mathrm{F}$ used by different authors to describe such many-body renormalized observables
of graphene as electron transition energies or quantum capacitance fall in the range $(0.8-1)\times10^6\,\mbox{m/s}$.

\begin{table}[b]
\caption{\label{Table1}Dielectric constants of surrounding medium $\varepsilon$, which provide the best least-square
fittings of the experimental data from Ref.~\cite{Luican} at several selected $v_\mathrm{F}$. The fittings are carried
out in four screening models, described in Sec.~\ref{sec2}.}
\centering
\begin{tabular}{ccccc}
\hline%
\hline%
&Unscreened&Screened&Self-cons.&Self-cons.\\
$v_\mathrm{F},$&interaction&interaction&screening&screening\\
$10^6\,\mbox{m/s}$&$r_\mathrm{s}=0$&$r_\mathrm{s}=\frac{e^2}{\varepsilon\hbar v_\mathrm{F}}$&$
r_\mathrm{s}=\frac{e^2}{\varepsilon\hbar\langle v_\mathrm{F}^*\rangle}$&
$r_\mathrm{s}=\frac{e^2}{\varepsilon\hbar v_\mathrm{F}^*(n_\mathrm{e})}$\\
\hline%
0.8\hphantom{0}&\hphantom{1}5.83&$<0.01$&1.40&1.46\\
0.85&\hphantom{1}6.54&0.35&2.01&1.95\\
0.9\hphantom{0}&\hphantom{1}7.45&1.36&2.79&2.60\\
1\hphantom{.00}&10.31&4.31&5.33&4.70\\
\hline%
\hline%
\end{tabular}
\end{table}

Table~\ref{Table1} shows the results of adjusting $\varepsilon$ at several selected values of $v_\mathrm{F}$ in the
four models, made in order to achieve the best least-square fits of the experimental points
$v_\mathrm{F}^*(n_\mathrm{e})$. In the absence of the screening, we have obtained the overestimated values of
$\varepsilon$ because higher external dielectric constants are required to simulate the screening of the interaction
caused by both surrounding medium and graphene electrons. On the contrary, the non-self-consistent static screening
requires very low $\varepsilon$ (sometimes even unphysically smaller than 1), which indicates an overestimation of the
actual screening in this model. Finally, the two self-consistent screening models provide more realistic values of
$\varepsilon$. To determine $r_\mathrm{s}$ in the 3rd model, we take the experimental average $\langle
v_\mathrm{F}^*\rangle=1.253\times10^6\,\mbox{m/s}$.

Similarly to our earlier work \cite{Sokolik} and to \cite{Yu}, we find the bare Fermi velocity
$v_\mathrm{F}=0.85\times10^6\,\mbox{m/s}$ as providing the best agreement of the many-body theory with the experiments.
In the most accurate 3rd and 4th screening models, we get $\varepsilon\approx2$ which is close to the experimental
effective dielectric constant $~2.5$ arising when graphene is supported by the $\mathrm{SiO}_2$ substrate from the
bottom.

Fig.~\ref{Fig3} shows the best fits at this $v_\mathrm{F}$ in four screening models, which correspond to the entries in
the second row of Table~\ref{Table1}. Due to noticeable scattering of the experimental points (see the inset), it is
better to compare the calculations not with the points themselves, but with their smooth logarithmic trend
(\ref{Luican_trend}). The 3rd and 4th self-consistent screening models, which use $r_\mathrm{s}$ determined from either
experimental or from iteratively calculated $v_\mathrm{F}^*$, are the best in reproducing the experimental trend in the
whole range of $n_\mathrm{e}$ with realistic $\varepsilon$.

\begin{figure}[t]
\begin{center}
\resizebox{0.9\columnwidth}{!}{\includegraphics{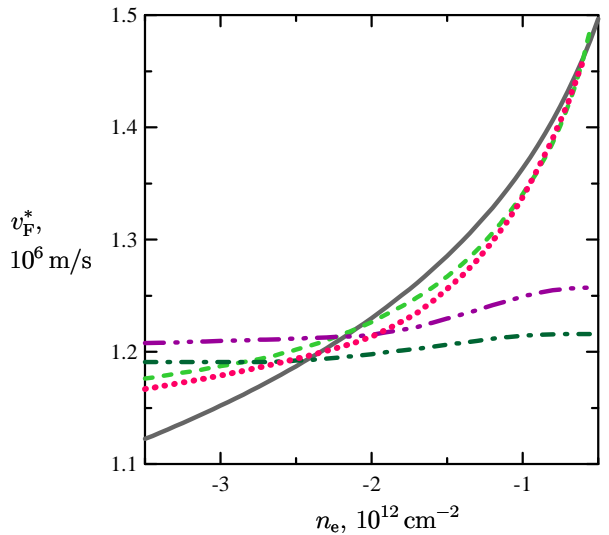}}
\end{center}
\caption{\label{Fig4}The same as Fig.~\ref{Fig3}, but only the models of the screened interaction (short-dashed line)
and the self-consistently screened interaction with the iteratively calculated varying $r_\mathrm{s}$ (dotted line) are
retained. Calculations in the same approximations but with the screening corresponding to the constant filling factor
$\nu=-14$ are shown by, respectively, dash-dotted and dash-double-dotted lines. The logarithmic trend
(\ref{Luican_trend}) of the experimental Fermi velocity growth is shown by think solid line.}
\end{figure}

As seen, the unscreened interaction does not describe the rapid growth of $v_\mathrm{F}^*$ on approach to the charge
neutrality point even at the optimal $\varepsilon$. Therefore the conventional first-order logarithmic renormalization
of the Fermi velocity \cite{Gonzalez} fails to reproduce the experiment \cite{Luican} with realistic model parameters
$v_\mathrm{F}$, $\varepsilon$. The screening resolves this problem because at smaller $|n_\mathrm{e}|$ the
polarizability of graphene decreases and thus the upward renormalization of the Fermi velocity becomes stronger. To
illustrate this point, we compare in Fig.~\ref{Fig4} the calculations in the 2nd and 4th screening models, which are
carried out, on the one hand, with the screening correctly dependent on $n_\mathrm{e}$, and, on the other hand, with a
constant polarizability ``frozen'' at the typical filling factor $\nu=-14$. We see that the latter calculations, which
take into account only changes of the occupation numbers $f_{n'}$ in (\ref{Sigma}) on varying the carrier density, does
not reproduce the rapid growth of $v_\mathrm{F}^*$. So the dependence of $\Pi(q,0)$ on $n_\mathrm{e}$ is crucial to
achieve agreement with the experiment.

\begin{table}[b]
\caption{\label{Table2}Dielectric constants of surrounding medium $\varepsilon$, which provide the best least-square
fittings of the experimental data from Ref.~\cite{Chae} at several selected $v_\mathrm{F}$. The fittings are carried
out in four screening models, described in Sec.~\ref{sec2}.} \centering
\begin{tabular}{ccccc}
\hline%
\hline%
&Unscreened&Screened&Self-cons.&Self-cons.\\
$v_\mathrm{F},$&interaction&interaction&screening&screening\\
$10^6\,\mbox{m/s}$&$r_\mathrm{s}=0$&$r_\mathrm{s}=\frac{e^2}{\varepsilon\hbar v_\mathrm{F}}$&$
r_\mathrm{s}=\frac{e^2}{\varepsilon\hbar\langle v_\mathrm{F}^*\rangle}$&
$r_\mathrm{s}=\frac{e^2}{\varepsilon\hbar v_\mathrm{F}^*(n_\mathrm{e})}$\\
\hline%
0.8\hphantom{0}&\hphantom{1}7.57&\hphantom{1}1.50&\hphantom{1}2.93&\hphantom{1}2.94\\
0.85&\hphantom{1}9.05&\hphantom{1}3.08&\hphantom{1}4.27&\hphantom{1}4.26\\
0.9\hphantom{0}&11.23&\hphantom{1}5.32&\hphantom{1}6.27&\hphantom{1}6.25\\
1\hphantom{.00}&21.76&15.58&16.12&16.09\\
\hline%
\hline%
\end{tabular}
\end{table}

\begin{figure}[t]
\begin{center}
\resizebox{0.9\columnwidth}{!}{\includegraphics{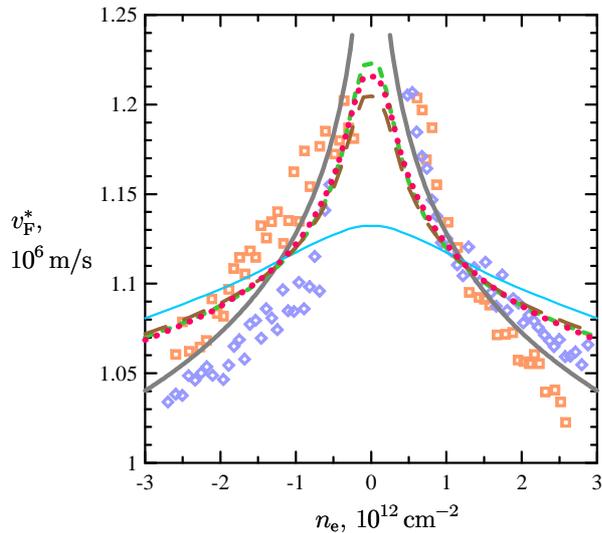}}
\end{center}
\caption{\label{Fig5}The best fits to the experimental points \cite{Chae} (squares and diamonds, corresponding to
electron and hole puddles) on $v_\mathrm{F}^*$ versus carrier density $n_\mathrm{e}$ at
$v_\mathrm{F}=0.85\times10^6\,\mbox{m/s}$. The designations of the curves are the same as in Fig.~\ref{Fig3}. The
logarithmic trend (\ref{Chae_trend}) of the experimental data is shown by thick solid line.}
\end{figure}

In the second considered experiment \cite{Chae}, graphene on a hexagonal boron nitride substrate was studied and
energies of larger number of Landau levels, typically $n=-8,-7,\ldots,+7,+8$, at magnetic fields $B=2$ and
$5\,\mbox{T}$ were measured. Two sets of experimental points $v_\mathrm{F}^*(n_\mathrm{e})$ were obtained in the range
$-3\times10^{12}\,\mbox{cm}^{-2}<n_\mathrm{e}<3\times10^{12}\,\mbox{cm}^{-2}$ in two locations inside electron and hole
puddles. These points follow the logarithmic trend
\begin{eqnarray}
\frac{v_\mathrm{F}^*(n_\mathrm{e})}{10^6\,\mbox{m/s}}=
\left(1.128-0.0799\ln\frac{|n_\mathrm{e}|}{10^{12}\,\mbox{cm}^{-2}}\right).\label{Chae_trend}
\end{eqnarray}
We have taken in our calculations 34 combinations of $n=-8,-7,\ldots,+7,+8$, and $B=2,5\,\mbox{T}$ to evaluate
$v_\mathrm{F}^*(n_\mathrm{e})$ in the same range.

Similarly to the analysis of the previous experiment, we have carried out the least-square fitting by adjusting
$\varepsilon$, and the results are given in Table~\ref{Table2}. In the 3rd model we take the average experimental Fermi
velocity $\langle v_\mathrm{F}^*\rangle=1.105\times10^6\,\mbox{m/s}$. Here we see the same regularities as in the
previous analysis, but the optimal dielectric constants turn out to be larger, because here the measured renormalized
Fermi velocities are generally lower than in \cite{Luican} due to a stronger screening from the substrate. The best
results with $v_\mathrm{F}=0.85\times10^6\,\mbox{m/s}$ in the self-consistent screening models provide
$\varepsilon\approx4-4.5$ which are close to the experimental $\varepsilon=3.15$ \cite{Chae} corresponding to graphene
on the hexagonal boron nitride substrate. The best fits at $v_\mathrm{F}=0.85\times10^6\,\mbox{m/s}$ in the four
screening models, corresponding to the entries in the second row of Table~\ref{Table2}, are shown in Fig.~\ref{Fig5}.

\begin{figure}[t]
\begin{center}
\resizebox{0.9\columnwidth}{!}{\includegraphics{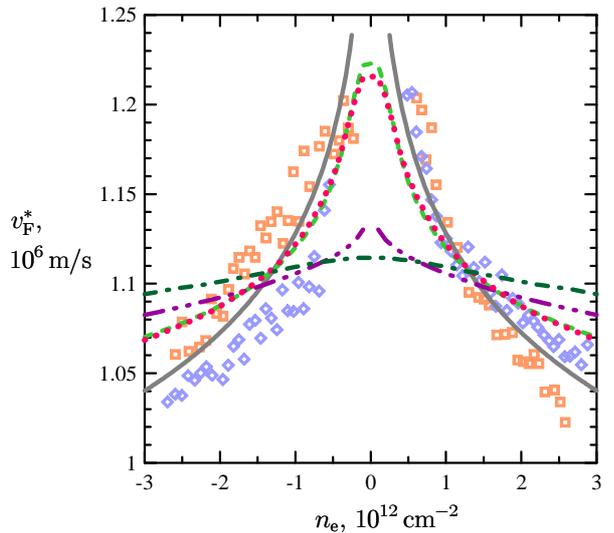}}
\end{center}
\caption{\label{Fig6}The same as Fig.~\ref{Fig4}, but for experimental points of \cite{Chae} with the logarithmic trend
(\ref{Chae_trend}).}
\end{figure}

Again, we see that the unscreened interaction cannot provide a sufficiently rapid growth of $v_\mathrm{F}^*$ near the
charge neutrality point. To illustrate the effect of the screening weakening at small carrier densities, we compare in
Fig.~\ref{Fig6} the calculations with the polarizability dependent on $n_\mathrm{e}$ and with that ``frozen'' at the
filling factor $\nu=-14$. Similarly to Fig.~\ref{Fig3}, this effect turns out to be crucial to correctly reproduce the
experimental data of \cite{Chae}.

\section{Conclusions}\label{sec5}

Using the statically screened Coulomb interaction in the random-phase approximation, we carried out the many-body
calculations of renormalized Landau level energies $E_n^*$ in graphene in quantizing magnetic field. Fitting a set of
$E_n^*$ at different $(n,B)$ by the formula (\ref{E_n_ren}), we evaluated the renormalized Fermi velocity
$v_\mathrm{F}^*$ as a function of the carrier density $n_\mathrm{e}$ and compared our calculations with the two
scanning tunneling spectroscopy experiments \cite{Luican,Chae}. We achieved good agreement using the bare Fermi
velocity $v_\mathrm{F}=0.85\times10^6\,\mbox{m/s}$ and the realistic values of the dielectric constant $\varepsilon$ as
the model parameters. The same value of $v_\mathrm{F}$ proved to be optimal in our previous work \cite{Sokolik} where
the cyclotron resonance and magneto-Raman scattering experimental data were described.

Our analysis allows us to draw the following main conclusions. First, the characterization of several renormalized
energies $E_n^*$ of different Landau levels $n$ at different magnetic fields $B$ at once with a single renormalized
Fermi velocity $v_\mathrm{F}^*$, assumed in many experimental works
\cite{Li1,Li2,Luican_SSC,Miller,Song,Luican,Chae,Yin2}, generally works very well. However there exist small, but
systematic deviations from the formula (\ref{E_n_ren}), both predicted by the theory and found in some experiments. In
particular, the logarithmic decrease of $v_\mathrm{F}^*$ with increasing $B$ was measured in magneto-Raman scattering
\cite{Faugeras1,Faugeras2}. Small deviations from the linear fits (\ref{E_n_ren}), similar to those presented in
Fig.~\ref{Fig1} can be noted in the results of several other experimental works \cite{Miller,Chen,Yin2} and thus can
possibly be attributed to many-body effects.

Second, the screening of Coulomb interaction is necessary to achieve a quantitative agreement between the theory and
the experiments with realistic values of $v_\mathrm{F}$ and $\varepsilon$. Besides, the calculations without the
screening fail to reproduce the rapid growth of $v_\mathrm{F}^*$ on approach to the charge neutrality point reported in
\cite{Luican,Chae}. It can be reproduced only with taking into account that the screening becomes weaker when the
carrier density decreases.

Third, the self-consistent weakening of the screening due to interaction-induced enlargement of virtual electron
transition energies is also important to obtain quantitatively correct theoretical results. In our previous work
\cite{Sokolik}, we made the same conclusion based on the analysis of the other experiments \cite{Jiang1,Faugeras1}. In
this work, we additionally demonstrate that the rapidly converging iterative self-consistent calculations of the Landau
level energies and polarizability are possible in the case of graphene in a magnetic field.

Finally, we note that although the logarithmic growth of the quasiparticle velocity $v_\mathrm{F}^*$, predicted when
the particle momentum approaches the Dirac point $k\rightarrow0$ in undoped graphene \cite{Kotov,Gonzalez,CastroNeto},
is similar to the growth of $v_\mathrm{F}^*$ in doped graphene when the carrier density decreases
$n_\mathrm{e}\rightarrow0$, their physical origins are not the same. In the second case, the key role of the
carrier-density dependent screening should be taken into account.

\section*{Acknowledgments}
The authors are grateful to Andrey D. Zabolotskiy for helpful discussions. The work of Y.E.L. was supported by the
Program for Basic Research of the National Research University Higher School of Economics, and the work of A.A.S. was
supported by the Foundation for the advancement of theoretical physics ``BASIS''.

\end{document}